\documentclass[conference]{IEEEtran}
\IEEEoverridecommandlockouts

\usepackage[utf8]{inputenc}
\usepackage[whole]{bxcjkjatype}  %

\usepackage{graphicx}
\usepackage{xcolor}
\usepackage[caption=false,font=footnotesize]{subfig}
\usepackage{url}
\usepackage{cite}

\usepackage{array, amsmath, amssymb, amsfonts, bm}
\usepackage{mathrsfs}
\usepackage{multirow, booktabs}
\usepackage{algorithm, algpseudocode}
\usepackage{xargs}

\usepackage{caption}
\usepackage{comment}
\usepackage{kantlipsum}
\usepackage{todonotes}
\usepackage{microtype}

\def\BibTeX{{\rm B\kern-.05em{\sc i\kern-.025em b}\kern-.08em
    T\kern-.1667em\lower.7ex\hbox{E}\kern-.125emX}}

\newcommand{\diag}{\mathrm{diag}}

\newcommand{\setR}{\mathbb{R}}
\newcommand{\setRp}{\mathbb{R}_+}
\newcommand{\setC}{\mathbb{C}}
\newcommand{\setSp}{\mathbb{S}_+}

\newcommand{\eye}{{\bf I}}

\newcommand{\T}{\mathsf{T}}
\newcommand{\adj}{\mathsf{H}}

\newcommand{\distnormal}[2]{\mathcal{N}\left({#1}, {#2}\right)}
\newcommand{\distcmpnormal}[2]{\mathcal{N}_{\mathbb{C}}\left({#1}, {#2}\right)}

\newcommand{\E}{\mathbb{E}}
\newcommand{\KL}{\mathcal{D}_\mathrm{KL}}

\NewDocumentCommand\newletter{m m o m m}{%
\NewDocumentCommand#1{s t@ o}{%
\IfBooleanTF{##1}{\mathbf{\MakeUppercase{#2}}\IfValueT{#3}{^{#3}}}{%
\IfBooleanTF{##2}{\mathbf{#2}\IfValueT{#3}{^{#3}}_{\IfValueTF{##3}{##3}{#5}}}{%
{#2}\IfValueT{#3}{^{#3}}_{\IfValueTF{##3}{##3}{#4}}%
}}}}

\NewDocumentCommand\newletterbm{m m o m m}{%
\NewDocumentCommand#1{s t@ o}{%
\IfBooleanTF{##1}{\bm{\MakeUppercase{#2}}\IfValueT{#3}{^{#3}}}{%
\IfBooleanTF{##2}{\bm{#2}\IfValueT{#3}{^{#3}}_{\IfValueTF{##3}{##3}{#5}}}{%
{#2}\IfValueT{#3}{^{#3}}_{\IfValueTF{##3}{##3}{#4}}%
}}}}

\newletter{\x}{x}{ftm}{ft}

\newletterbm{\psd}{\lambda}{nft}{ft}

\newcommand{\scm}[1][nf]{\mathbf{H}_{#1}}

\newcommand{\Y}[1][:ft]{\mathbf{Y}_{#1}}
\newcommand{\Q}[1][f]{\mathbf{Q}_{#1}}
\newletter{\q}{q}{fmm}{fm}
\newletter{\xt}{\tilde{x}}{ftm}{ft}
\newletter{\yt}{\tilde{y}}{ftm}{ft}

\newletter{\z}{z}{ntd}{nt}
\newletter{\zs}{z}[*]{ntd}{nt}

\newletter{\g}{w}{nm}{n}
\newcommand{\G}[1][nf]{\mathbf{W}_{#1}}

\newcommand{\sv}[1][nf]{\mathbf{a}_{#1}}

\newcommand{\dec}[1][\theta,f]{g_{#1}}

\newletter{\src}{s}{nft}{nf}

\newcommand{\elbo}{\mathcal{L}}

\begin{document}

\setlength\abovedisplayskip{1.8mm}
\setlength\belowdisplayskip{1.8mm}

\hyphenpenalty=0
\linepenalty=999    

\title{Neural Fast Full-Rank Spatial Covariance Analysis \\ for Blind Source Separation\\
\thanks{This work was supported in part by the JST ACT-X under Grant JPMJAX200N and NEDO.}
}

\author{
    \IEEEauthorblockN{
        Yoshiaki Bando\IEEEauthorrefmark{1}\IEEEauthorrefmark{2}, %
        Yoshiki Masuyama\IEEEauthorrefmark{1}\IEEEauthorrefmark{3}, %
        Aditya Arie Nugraha\IEEEauthorrefmark{2}, %
        and Kazuyoshi Yoshii\IEEEauthorrefmark{2}\IEEEauthorrefmark{4}\vspace{.0\baselineskip}}
    \IEEEauthorblockA{
        \IEEEauthorrefmark{1}National Institute of Advanced Industrial Science and Technology, Japan}
    \IEEEauthorblockA{
        \IEEEauthorrefmark{2}Center for Advanced Intelligence Project (AIP), RIKEN, Tokyo, Japan}
    \IEEEauthorblockA{
        \IEEEauthorrefmark{3}Department of Computer Science, Tokyo Metropolitan University, Japan}
    \IEEEauthorblockA{
        \IEEEauthorrefmark{4}Graduate School of Informatics, Kyoto University, Japan}
}

\bstctlcite{IEEE:BSTcontrol}

\maketitle
\begin{abstract}
This paper describes an efficient unsupervised learning method for a neural source separation model
 that utilizes a probabilistic generative model 
 of observed multichannel mixtures
 proposed for blind source separation (BSS).
For this purpose,
 amortized variational inference (AVI) 
 has been used for directly solving the inverse problem
 of BSS with full-rank spatial covariance analysis (FCA).
Although this unsupervised technique called neural FCA 
 is in principle free from the domain mismatch problem,
 it is computationally demanding 
 due to the full rankness of the spatial model
 in exchange for robustness against relatively short reverberations.
To reduce the model complexity without sacrificing performance,
 we propose neural FastFCA 
 based on the jointly-diagonalizable yet full-rank spatial model.
Our neural separation model introduced for AVI alternately performs
 neural network blocks and
 single steps
 of an efficient iterative algorithm called iterative source steering.
This alternating architecture enables the separation model to 
 quickly separate the mixture spectrogram by leveraging both the deep neural network and the multichannel optimization algorithm.
The training objective with AVI is 
 derived to maximize the marginalized likelihood of the observed mixtures.
The experiment using mixture signals of two to four sound sources 
 shows that neural FastFCA outperforms conventional BSS methods 
 and reduces the computational time to about 2\,\% of that for the neural FCA.
\end{abstract}

\section{Introduction}

Sound source separation forms the basis of various machine listening systems including distant speech recognition~\cite{watanabe2020chime,du2020ustc,subramanian2020far} and sound event detection~\cite{scheibler2022sound,turpault2020improving}.
Neural source separation has achieved excellent performance thanks to the expression power of deep neural networks (DNNs) trained with a large number of pairs of mixture signals and their corresponding source signals~\cite{zhu2021multi,tzinis2020sudo,luo2019convtasnet}.
However, such supervised training suffers from domain mismatch and a lack of source signals in target environments.
As a promising alternative, blind source separation (BSS)~\cite{ono2011stable,sawada2013multichannel,sekiguchi2020fast} has thus been investigated to work with little prior information about the sources and microphones.

Modern BSS methods
 are based on probabilistic generative models of multichannel mixture signals~\cite{sekiguchi2020fast,sawada2013multichannel,
 ono2011stable,ozerov2009multichannel}.
Such a probabilistic model
 consists of a source model representing the power spectral densities (PSDs) of the sources
 and a spatial model representing the spatial covariance matrices (SCMs) of the sources.
Multichannel non-negative matrix factorization (MNMF)~\cite{sawada2013multichannel}, 
 for example, is based on an NMF-based source model 
 assuming the low-rankness of the PSDs 
 and a full-rank spatial model assuming the full-rankness of the SCMs.
While the full-rank SCMs can deal with small source movements and reverberation, 
 their estimation is often unstable 
 and requires an expensive computational cost due to their too-high degrees of freedom.
FastMNMF~\cite{ito2019fastmnmf,sekiguchi2020fast} mitigates this problem by assuming the source SCMs to be jointly-diagonalizable (JD).
The JD SCM is also full-rank but is represented by a weighted sum of rank-1 SCMs common to all the sources.
This constraint is reported to efficiently reduce the computational cost and improve the separation performance compared to the original MNMF~\cite{sekiguchi2020fast}.

To represent complex structures of source spectra, neural source models using variational autoencoders (VAEs)~\cite{kingma2013auto} have been proposed~\cite{sekiguchi2019semi,leglaive2019semi,kameoka2018semiblind,bando2021neural}.
For example, a multichannel VAE (MVAE)~\cite{kameoka2018semiblind} replaces the source model of MNMF with the decoder of a VAE pre-trained on isolated source signals.
This source model can also be trained only with mixture signals by neural full-rank spatial covariance analysis (FCA)~\cite{bando2021neural}.
Neural FCA trains the source generative model (decoder) by introducing an inference (encoder) model that estimates latent features of the source model from a multichannel mixture.
The decoder and encoder models are jointly trained to maximize the likelihood of the MVAE for training data of multichannel mixtures.
The neural FCA was reported to perform on par with the supervised MVAE for speech separation~\cite{bando2021neural}.

\begin{figure}[t]
  \centering
  \includegraphics[width=\hsize]{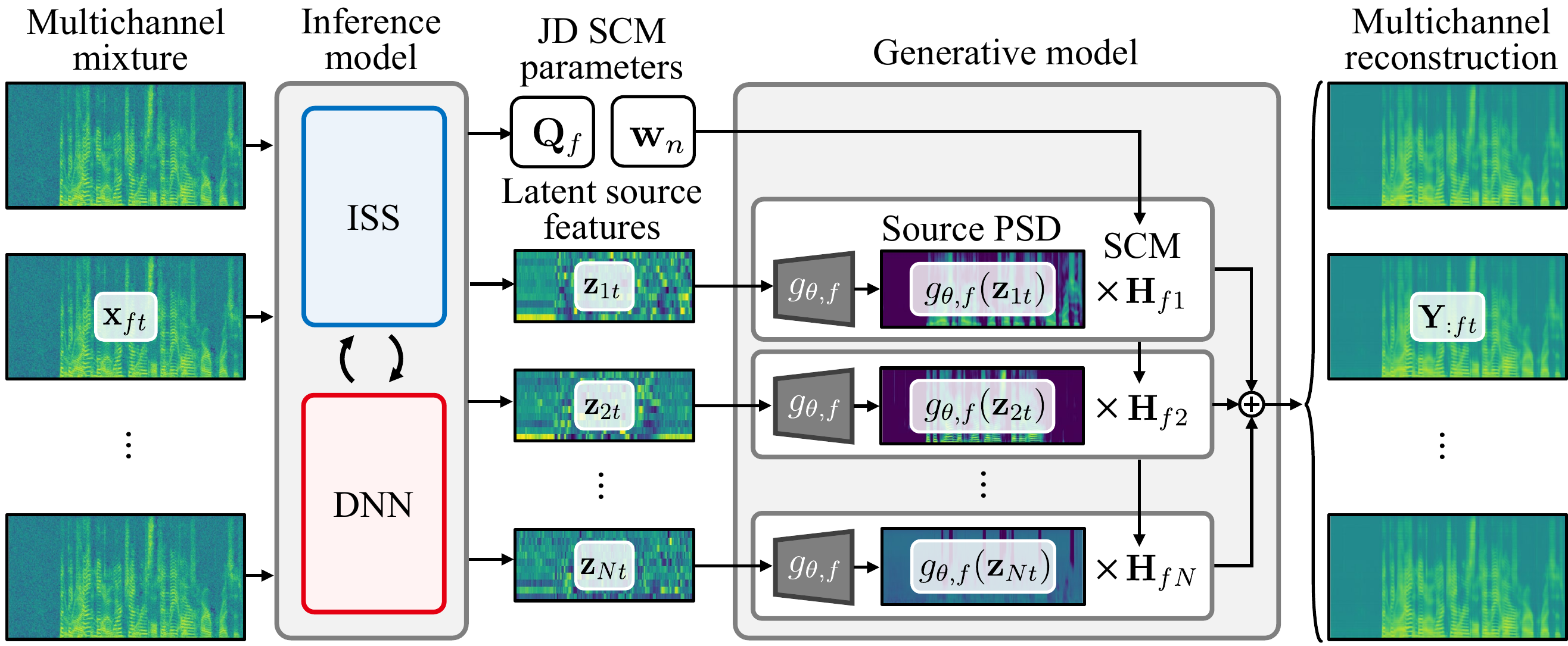}
  \caption{The overview of the proposed neural FastFCA.}
  \label{fig:overview}
\end{figure}

In this paper, we propose a BSS method called neural FastFCA based on the integration of the JD spatial model and the neural source model (Fig.~\ref{fig:overview}).
The original neural FCA estimates the full-rank SCMs by an expectation-maximization (EM) algorithm, which requires a high computational cost.
In contrast, we assume the JD spatial model and extend the inference model to estimate the JD SCMs quickly in the network.
Specifically, we introduce a network building block that diagonalizes an observed mixture by an efficient algorithm called iterative source steering (ISS)~\cite{Scheibler2021,Saijo2022spatial}.
We alternately stack the ISS-based diagonalization blocks and DNN blocks such that the intermediate diagonalization (quasi-separation) results can be used to estimate the latent source features.
The networks are jointly trained to separate unseen mixture signals in an unsupervised manner.

The main contribution of this study is to integrate the state-of-the-art BSS techniques of the JD spatial model~\cite{sekiguchi2020fast}, the neural source model~\cite{bando2021neural}, and the ISS-based inference model~\cite{Scheibler2021}.
This combination enables the proposed method to train the inference (separation) model and the neural source model in an unsupervised manner to achieve high separation performance and a small computational cost.
The experimental results with simulated mixture signals of two to four speech sources demonstrate that our blind method outperforms conventional BSS methods.
In addition, our method reduces the computational cost to about 2\,\% of that for the original neural FCA.

\section{Background}
This section first briefly overviews the existing BSS methods and then introduces a neural BSS method called neural FCA.

\subsection{Blind source separation}
BSS methods typically assume that an $M$-channel mixture signal $\x@ \in \setC^M$ is a sum of $N$ source signals $\src \in \setC$:
\begin{align}
  \x@ = \sum_{n=1}^N \sv \src,
\end{align}
where $t=1,\ldots, T$ and $f=1,\ldots,F$ are time and frequency indices, respectively, and $\sv \in \setC^M$ is the steering vector for source $n$.
Each source signal $\src$ is then assumed to follow a zero-mean complex Gaussian distribution as follows:
\begin{align}
  \src \sim \distcmpnormal{0}{\psd},
\end{align}
where $\psd \in \setRp$ represents the PSD of source $n$.
By marginalizing the source signal $\src$, the following multivariate Gaussian likelihood is obtained:
\begin{align}
  \x@ \sim \distcmpnormal{\bm{0}}{\sum_{n=1}^N \psd \scm}, \label{eq:lgm}
\end{align}
where $\scm = \sv \sv^\adj \in \setSp^{M\times M}$ is an SCM for source $n$ at frequency $f$.
BSS is performed by estimating the $\psd$ and $\scm$ to maximize this likelihood with sufficient assumptions to effectively restrict the model's redundant flexibility.
Independent low-rank matrix analysis (ILRMA)~\cite{kitamura2016determined}, for example, assumes $\psd$ to be low-rank for solving frequency permutation ambiguity.
MNMF~\cite{ozerov2009multichannel} replaces $\scm$ with a full-rank SCM for allowing small source movements and reverberations.

The computational cost for estimating the full-rank SCMs can be efficiently reduced by using the JD SCMs~\cite{ito2019fastmnmf,sekiguchi2020fast}.
Specifically, this formulation represents SCMs $\scm$ by a diagonalizer $\Q \in \setC^{M\times M}$ common for all the sources and diagonal elements $\g@ \in \setRp^M$ for each source as follows:
\begin{align}
  \scm = \Q^{-1} \diag(\g@) \Q^{-\adj}. \label{eq:jd}
\end{align}
The diagonalizer $\Q$ is optimized to maximize Eq.~\eqref{eq:lgm} with an iterative projection~\cite{ono2011stable} or ISS~\cite{scheibler2020fast} algorithm, 
 and $\g@$ is optimized with a multiplicative update rule~\cite{sekiguchi2020fast}.
FastMNMF combines this JD spatial model with a low-rank source model and has been reported to perform better than MNMF while working at a similar computational cost to ILRMA~\cite{sekiguchi2020fast}.

\subsection{Neural full-rank spatial covariance analysis}
A powerful way to represent source signals is to utilize a DNN that can precisely capture their complex spectra~\cite{sekiguchi2019semi,leglaive2019semi,kameoka2018semiblind}.
The deep spectral model~\cite{kameoka2018semiblind,sekiguchi2019semi,leglaive2019semi} assumes that the PSD $\psd$ is generated by a latent source feature $\z@ \in \setR^D$ and a non-linear function (\textit{i.e.}, DNN) $\dec: \setR^D\rightarrow \setRp$ as follows:
\begin{align}
  \psd = \dec(\z@), \label{eq:dsp}
\end{align}
where $\theta$ is a set of the network parameters of $\dec$.
The latent source feature $\z@$ is typically assumed to follow the standard Gaussian distribution:
\begin{align}
  \z@ \sim \distnormal{\bm{0}}{\eye}, \label{eq:pz}
\end{align}
and is supposed to represent the features of source spectra, such as pitches and envelopes.

The neural FCA~\cite{bando2021neural} trains the neural source model $\dec$ as a decoder of a VAE by introducing an inference (encoder) model with network parameters $\phi$ to estimate the posterior distribution $q_\phi(\z@ \mid \x*)$ from an observed mixture $\x* \triangleq \{ \x@ \}_{f,t=1}^{F, T}$.
Let $\z* \triangleq \{\z@\}_{n,t=1}^{N,T}$ and $\scm[] \triangleq \{\scm\}_{f,n=1}^{F,N}$ be the sets of latent features and SCMs, respectively.
This method assumes the generative model of multichannel mixtures with Eqs.~\eqref{eq:lgm}, \eqref{eq:dsp}, and \eqref{eq:pz}.
Based on this generative model, the encoder and decoder are jointly trained in an unsupervised manner to maximize the following evidence lower bound (ELBO):
\begin{align}
  \elbo = \E_{q_\phi}[\log p_\theta(\x* \mid \z*, \scm[])] - \KL[q_\phi(\z*\mid \x*) \mid p(\z*)],
\end{align}
where $\E_{q_\phi}[\cdot]$ and $\KL[ \cdot \mid \cdot ]$ are the expectation by the posterior $q_\phi$ and the Kullback-Leibler (KL) divergence, respectively.
The network parameters $\theta$ and $\phi$ are optimized by stochastic gradient ascent~\cite{kingma2014adam}, and the SCMs $\scm$ are optimized at each network update with an EM algorithm~\cite{duong2010underdetermined}.
The maximization of the ELBO corresponds to the maximization of the log-marginal likelihood $p(\x*\mid\scm[])$, and can be considered as BSS performed for the training mixture signals.

Once the networks are optimized, they are used to separate unseen mixture signals.
Neural FCA has been reported to perform better than existing BSS methods including FastMNMF and on par with the supervised MVAE in speech separation~\cite{bando2021neural}.
This method, however, requires a high computational cost for estimating the SCMs.
In addition, the separation performance is limited because the inference network does not utilize the intermediate separation results, which are usually utilized in the conventional BSS methods during their iterative algorithms.

\section{Neural Fast Full-Rank Spatial \\ Covariance Analysis}
We extend the original neural FCA with the JD spatial model to reduce the computational cost without sacrificing performance.
In addition, we introduce an inference model based on the tight integration of ISS-based blocks and DNN blocks for quickly separating multichannel mixture signals.

\subsection{Generative model of multichannel mixture signals}
Our method called neural FastFCA is based on the JD spatial model of Eqs.~\eqref{eq:lgm} and \eqref{eq:jd} and the neural source model of Eqs.~\eqref{eq:dsp} and \eqref{eq:pz}.
The resulting generative model is as follows:
\begin{align}
  \hspace{-2.0mm}\x@ \sim \distcmpnormal{\bm 0}{\Q^{-1}\left\{ \sum_{n=1}^N \dec(\z@) \diag(\g@)\right\} \Q^{-\adj}}.
\end{align}

\subsection{Inference model}
\begin{figure}[t]
  \centering
  \includegraphics[width=\hsize]{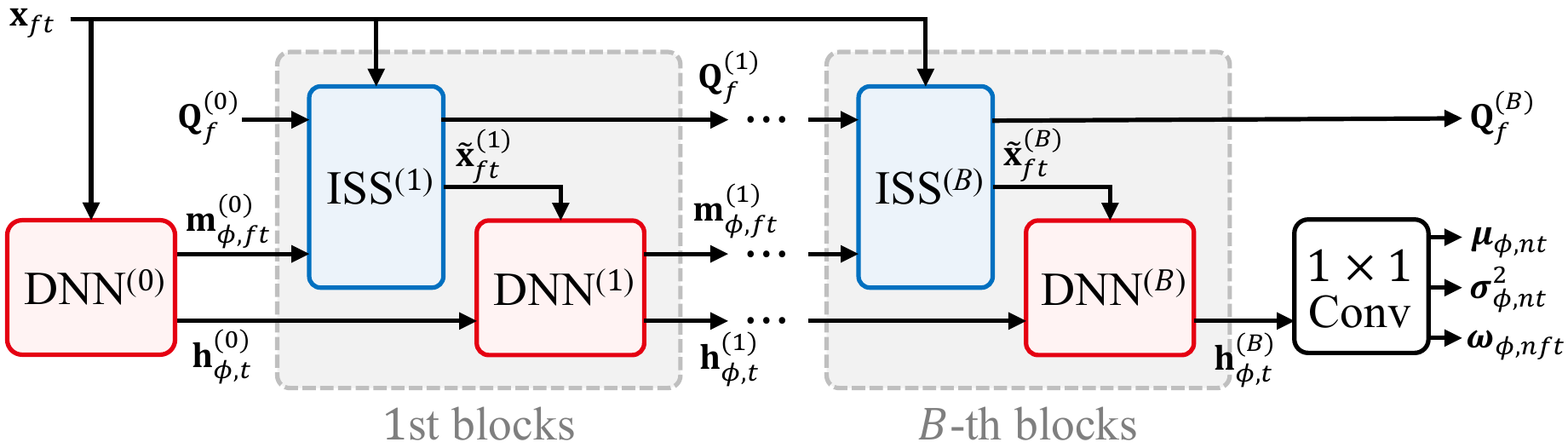}
  \caption{The block diagram of the inference model.}
  \label{fig:inference}
\end{figure}
The inference (separation) model of neural FastFCA estimates SCM parameters $\Q[] \triangleq \{ \Q \}_{f=1}^{F}$ and $\G[] \triangleq \{ \g@ \}_{n=1}^{N}$ as well as the posterior distribution $q_\phi(\z*\mid \x*)$ from an observed mixture $\x*$.
We utilize both the DNNs and multichannel optimization techniques~\cite{Scheibler2021} for quick source separation.
Specifically, as shown in Fig.~\ref{fig:inference}, the inference network estimates the parameters by alternately performing $B+1$ DNN blocks and $B$ ISS blocks within the network.
The $b$-th ISS block~\cite{Scheibler2021} updates the diagonalizer $\Q \triangleq [\q@[f1], \ldots \q@[fM]]^\adj$ by iterating the following ISS update rule for $m=1,\ldots, M$:
\begin{align}
  \Q &\leftarrow \Q - [v_{fm1}, \ldots, v_{fmM}]^\T \q@[fm]^\adj, \\
  v_{fmm'} &= \begin{cases}
    \frac{\q@[fm']^\adj \mathbf{U}_{fm'}\q@[fm]}{\q@[fm]^\adj \mathbf{U}_{fm'}\q@[fm]} & \text{(if $m' \neq m$)} \\
    1 - (\q@[fm]^\adj \mathbf{U}_{fm}\q@[fm])^{-\frac{1}{2}}                        & \text{(if $m' = m$)}
  \end{cases}
\end{align}
where $\mathbf{U}_{fm} \in \setSp^{M\times M}$ is an auxiliary SCM.
This SCM is calculated by the mixture $\x@$ and a TF mask $m^{(b-1)}_{\phi,ftm'} \in [0, 1]$ predicted by the $(b-1)$-th DNN block:
\begin{align}
  \mathbf{U}_{fm'} &= \frac{1}{T} \sum_{t=1}^T m^{(b-1)}_{\phi,ftm'} \cdot \x@ \x@^\adj.
\end{align}
Let $\Q^{(b)}$ be the output of the $b$-th ISS block.
This update rule converges to the maximum likelihood estimate of $\Q$ by using an appropriate mask $m^{(b)}_{\phi,ftm'}$ and initial value $\Q^{(0)}$~\cite{scheibler2020fast}.
The $b$-th DNN block, on the other hand, outputs the TF mask $m^{(b)}_{\phi,ftm'}$ and an internal feature $\mathbf{h}^{(b)}_{\phi,t}$ ($t=1,\ldots, T$) passed to the next block.
The input of the DNN block is an intermediate diagonalized (quasi-separated) spectrogram $\xt@^{(b)} \triangleq \Q^{(b)} \x@ \in \setC^{M}$ concatenated with internal features $\mathbf{h}^{(b-1)}_{\phi,t}$.

After performing $B$ ISS blocks and $B+1$ DNN blocks, the last internal feature $\mathbf{h}^{(B)}_{\phi,t}$ is converted to $q_\phi(\z* \mid \x*)$ and $\g*$ with an output (1$\times$1-convolution) layer.
The posterior distribution $q_\phi(\z* \mid \x*)$ is estimated as the following Gaussian distribution:
\begin{align}
  q_\phi(\z*\mid \x*) &\leftarrow \hspace{-2.5mm}\prod_{n,t,d=1}^{N,T,D}\hspace{-1.0mm} \distnormal{\z \bigm| \mu_{\phi,ntd}}{\sigma^2_{\phi,ntd}},
\end{align}
where $\mu_{\phi,ntd} \in \setR$ and $\sigma^2_{\phi,ntd} \in \setRp$ are the network outputs representing the mean and variance of $\z*$, respectively.
On the other hand, the diagonal elements $\g@[n]$ are estimated as a normalized average of frequency-wise estimates $\mathbf{w}'_{\phi,fn} \in \setRp^M$:
\begin{align}
  \g@[n]      &\leftarrow \frac{1}{F} \sum_{f=1}^F \frac{1}{\frac{1}{M}\lVert \mathbf{w}'_{\phi,fn} \rVert_1} \mathbf{w}'_{\phi,fn} \\
  \mathbf{w}'_{\phi,fn} &= \sum_{t=1}^T \bm{\omega}_{\phi,nft} \circ \left|\xt@^{(B)}\right|^{\circ2}
\end{align}
where $\bm{\omega}_{\phi,nft} \in [0, 1]^{M}$ is a network output to represent an $M$-channel TF mask, and $\circ$ and $\left|\cdot\right|^{\circ2}$ indicate the element-wise product and element-wise absolute square, respectively.

\begin{table*}[t]
    \centering
    \caption{Separation performance in SDR, PESQ, and STOI and elapsed time for separation in seconds.}
    \label{tab:scores}
    \vspace{-2mm}
    \begin{tabular}{l|c|c|ccc|ccc|ccc|ccc}
        \toprule
        \multirow{2}{*}{Method} & \# of & Elapsed & \multicolumn{3}{c|}{Average} & \multicolumn{3}{c|}{$K=2$} & \multicolumn{3}{c|}{$K=3$} & \multicolumn{3}{c}{$K=4$} \\
        & iters. & time & SDR & PESQ & STOI & SDR & PESQ & STOI & SDR & PESQ & STOI & SDR & PESQ & STOI \\
        \midrule
        MNMF                 & 200 & 2.07 &  7.5 & 1.49 & 0.76 & 13.0 & 1.93 & 0.85 &  8.3 & 1.47 & 0.79 & 3.9 & 1.26 & 0.69 \\
        ILRMA                & 200 & 1.36 &  7.0 & 1.43 & 0.76 & 13.2 & 1.83 & 0.86 &  7.7 & 1.39 & 0.79 & 3.2 & 1.24 & 0.69 \\
        FastMNMF            & 200 & 1.81 &  9.3 & 1.60 & 0.80 & 15.3 & 2.12 & 0.89 & 10.1 & 1.59 & 0.83 & 5.3 & 1.32 & 0.74 \\
        \midrule
        Neural FCA (fix $z$) & 200 & 2.67 &  8.9 & 1.71 & 0.79 & 15.2 & 2.28 & 0.89 & 10.1 & 1.75 & 0.83 & 4.5 & 1.36 & 0.71 \\
        Neural FCA           &   5 & 0.14 &  8.0 & 1.48 & 0.78 & 14.0 & 1.90 & 0.88 & 9.1 & 1.47 & 0.82 & 3.8 & 1.26 & 0.69 \\
        Neural FCA           &  10 & 0.26 &  8.6 & 1.53 & 0.79 & 14.6 & 1.98 & 0.89 & 9.7 & 1.52 & 0.83 & 4.3 & 1.28 & 0.70 \\
        Neural FCA           & 100 & 2.40 & 10.6 & 1.81 & 0.83 & 16.2 & 2.35 & 0.90 & 11.8 & 1.87 & 0.86 & 6.5 & 1.46 & 0.76 \\
        Neural FCA           & 200 & 4.77 & 11.1 & \bf 1.88 & 0.84 & 16.4 & \bf 2.41 & 0.90 & 12.2 & \bf 1.95 & 0.87 & 7.2 & \bf 1.52 & 0.78 \\
        \midrule
        Neural FastFCA (ours) &  -- & \bf 0.09 & \bf 11.6 & 1.85 & \bf 0.85 & \bf 17.4 & \bf 2.41 & \bf 0.91 & \bf 12.7 & 1.90 & \bf 0.88 & \bf 7.5 & 1.50 & \bf 0.79 \\
        \bottomrule
    \end{tabular}
  \end{table*}

\subsection{Amortized variational inference}\label{sec:training}
The generative and inference models are trained by using only multichannel mixture signals as an amortized variational inference~\cite{kingma2013auto}.
As in the original neural FCA, the training objective for each mixture signal is an ELBO as follows:
\begin{align}
  \elbo & = \E_{q_\phi}[\log p_\theta(\x* \mid \z*, \G[], \Q[])] -\KL[q_\phi(\z* \mid \x*) \mid p(\z*)], \nonumber
\end{align}
where $\Q[]$ denotes the network output $\Q[]^{(B)}$ for simplicity.
The KL term can be calculated in the same way as in~\cite{bando2021neural} and used to solve the frequency permutation ambiguity.
The first term of the ELBO, on the other hand, is calculated approximately from the inference results $\Q[]$, $\g*$, and $q_\phi(\z*\mid\x*)$ as follows:
\begin{align}
  \E[\log p_\theta(\x* \mid \z*, \G[], \Q[])] & \approx T \sum_{f=1}^F \log \left|\Q\Q^\adj\right| \nonumber                                                    \\
  & \hspace{-10mm}- \sum_{f,t,m=1}^{F, T, M} \left\{ \log \yt + \frac{|\xt|^2}{\yt} \right\},
\end{align}
where $\xt@ \triangleq [\xt[ft1],\ldots, \xt[ftM]]^T = \Q \x@ \in \setC^M$ is the diagonalized observation, and $\yt \triangleq \sum_{n=1}^N\g \dec(\zs@) \in \setRp$ is the mixture PSDs with a sample $\zs@ \sim q_\phi(\z@ \mid \x*)$.
Because all the operations for calculating the ELBO are differentiable, 
the networks are optimized by using stochastic gradient ascent.

\subsection{Source separation}
Once the generative and inference  models are trained, they are used to separate unseen mixture signals.
Specifically, the inference model first estimates the model parameters $\Q[]$, $\g*$, and $\hat{\mathbf{z}}_{ntd} = \bm{\mu}_{\phi,nt}(\x*)$.
The source signal $\hat{s}_{nft}$ is then estimated by a multichannel Wiener filter as follows:
\begin{align}
  \hat{s}_{nft} \leftarrow \mathbf{u}^\T \Y[nft]\Y[:ft]^{-1}\x@,
\end{align} %
where $\mathbf{u}$ is a one-hot vector representing a reference channel (the first channel in this paper), and $\Y = \sum_{n=1}^N\Y[nft]$ is the sum of source images $\Y[nft] = \dec(\hat{\mathbf{z}}_{ntd}) \Q^{-1}\diag(\g@)\Q^{-\adj}$.

\section{Experimental Evaluation}
The proposed method was evaluated with simulated mixture signals of various numbers of speech sources.

\subsection{Dataset}
We generated mixture signals of speech source signals by following the spatialized WSJ0-mix dataset~\cite{wang2018multi}.
Each mixture signal consisted of speech signals randomly selected from the WSJ0 English speech corpus~\cite{garofolo07wsj}.
We used the same subsets of speakers and utterances as in the WSJ0-mix dataset.
In contrast to the WSJ0-mix dataset, the number of source signals was randomly selected from $K \in \{2, 3, 4\}$.
A 6-channel microphone array ($M=6$) with random configuration is located randomly around the center of a room having random dimensions between 5\,m\,$\times$\,5\,m\,$\times$\,3\,m and 10\,m\,$\times$10\,m\,$\times$\,5\,m.
The sound sources were also randomly located while keeping the distance between each other more than 1\,m.
The reverberation time (RT$_{60}$) was randomly sampled between 200\,ms and 600\,ms, and the room impulse response for each source was simulated with the image method.
The speech signals were mixed at random powers uniformly chosen between $-2.5$\,dB and $+2.5$\,dB.
White diffuse noise with a signal-to-noise ratio of 30\,dB was added to each mixture signal as background noise.
We generated 20,000, 5,000, and 3,000 mixture signals for training, validation, and test sets, respectively.
All the mixtures were dereverberated by the weighted prediction error (WPE) method~\cite{yoshioka2012generalization}.

\subsection{Experimental condition}
The network architectures of the inference and generative models were experimentally determined as follows.
The inference model consisted of ISS and DNN blocks with $B=8$.
Each DNN block consisted of a U-Net-like architecture~\cite{tzinis2020sudo,bando2022weakly} having five 256-channel 1D-convolutional layers.
Each layer had a kernel size of $5$ and parametric rectified linear units (PReLUs).
The input feature of the $0$-th DNN block was the log-power spectrum and the inter-channel phase differences of an input mixture $\x@$.
That of the $b$-th ($b \geq 1$) DNN block was a concatenation of the internal feature $\mathbf{h}^{(b-1)}_{\phi,t}$ and a 512-dimensional vector converted from the log-power spectrum of $\xt@^{(b-1)}$ with a $1\times 1$-convolution layer.
We obtained the TF masks $m^{(b-1)}_{\phi,ftm}$ and $\bm{\omega}_{\phi,nft}$ with a sigmoid function and $\sigma^2_{\phi,ntd}$ with a softplus function.
The generative model $\dec$, on the other hand, consisted of three layers of 256-channel $1\times 1$-convolutional layers with PReLUs as in \cite{bando2021neural}.

We trained the inference and generative models with an Adam optimizer~\cite{kingma2014adam} for 200 epochs with a learning rate of $1.0\times 10^{-3}$.
The spectrograms were obtained using the short-time Fourier transform with a window length of 512 samples and a hop length of 128 samples.
The training was performed by splitting the spectrograms into 500-frame clips, and the batch size was set to 128 clips.
The dimension of the latent features was set to $D=50$.
The number of sound sources was set to $N=5$, assuming the maximum number of sources and diffuse noise ($4 + 1$).
We performed the cyclic annealing of the KL term in the ELBO~\cite{fu2019cyclical,bando2021neural}.
The diagonalizer was initialized to an identity matrix $\Q^{(0)} \leftarrow \eye$.
These hyperparameters were empirically determined using the validation set.

Our method was compared with existing BSS methods and the original neural FCA.
As BSS methods, we evaluated MNMF~\cite{sawada2013multichannel}, ILRMA~\cite{kitamura2016determined}, and FastMNMF~\cite{sekiguchi2020fast}.
The number of sources for MNMF and FastMNMF was set to $5$.
The numbers of bases and iterations for all the methods were set to $16$ and $200$, respectively.
The SCMs for MNMF were initialized by ILRMA.
The neural FCA~\cite{bando2021neural} had the same generative model $\dec$ as our method, and its inference model consisted of nine U-Net-like blocks to have the same number of blocks as our method.
We trained the neural FCA with the same hyperparameters as the proposed method.
The EM inference for SCMs $\scm$ was iterated $5$ times in the training phase following the literature~\cite{bando2021neural}.
At the test phase, the SCMs $\scm$ and latent features $\z@$ were updated to fit the observation with the EM rule and an Adam optimizer, respectively.
The learning rate for Adam was set to $0.2$.
We evaluated different numbers of iterations ($5$, $10$, $100$, and $200$ times) to assess how many iterations were needed for the conversion.

The performance was evaluated in terms of the signal-to-distortion ratio (SDR)~\cite{BSSEVAL} in dB, the perceptual evaluation of speech quality (PESQ)~\cite{rix2001perceptual}, and the short-term objective intelligibility (STOI)~\cite{taal2010short}.
They were evaluated with $K$ separated signals having the highest powers in the separation results.
We also measured the elapsed time for separating a $5$-second clip on an NVIDIA V100 accelerator with Intel Xeon Gold 6148 Processor.
To fully utilize the accelerator, the conventional BSS methods were implemented with CuPy 11.5.0, and the neural methods were implemented with PyTorch 1.13.1.

\subsection{Experimental results}

The separation performance for each number of sources $K \in \{2, 3, 4\}$ was summarized in TABLE~\ref{tab:scores}.
We first see that the original neural FCA required 200 times of updates for convergence.
Besides, the neural FCA deteriorated by fixing the latent features $\z@$ to the output of the inference model (``fix $z$" in the table).
This result indicates that the inference model failed to precisely estimate the latent source features only from the observed mixture.
In contrast, our neural FastFCA, which does not update the outputs of the inference model, outperformed the neural FCA in SDR and STOI and outperformed that without updates of $\z@$ in all the metrics.
In addition, our method reduced the computational time to about 2\,\% of that for the neural FCA (4.77 [s]).
The proposed method also clearly outperformed the conventional BSS methods of MNMF, ILRMA, and FastMNMF for all the conditions of $K \in \{2, 3, 4\}$.
We would also note that the neural FastFCA was trained successfully by using multichannel mixture signals to have different numbers of sources.
This result shows the promising possibility of our method to train a neural separation model in an unsupervised manner by specifying the maximum number of sources in the training mixtures.

\section{Conclusion}
This paper presented a neural BSS method called neural FastFCA based on the integration of the neural source model, JD spatial model, and ISS-based inference model.
Specifically, we extended the original neural FCA to have a JD full-rank spatial model to efficiently reduce the computational cost.
Our neural FastFCA also introduces an ISS-based inference model to improve the separation performance.
The experimental results with mixture signals having two to four sources showed that our neural FastFCA outperformed existing BSS methods.
In addition, the elapsed time for performing our method was reduced to 2\,\% of that for the original neural FCA.
Our future work includes further extending our method with various BSS techniques.
For example, the joint dereverberation and separation of moving sources is an important feature to computationally understand mixture signals recorded in indoor environments.
We will also investigate separating various kinds of sound sources in addition to speech signals.

\bibliographystyle{IEEEtran}
\bibliography{references}

\end{document}